# Magnetocaloric effect and magnetic cooling near a field-induced quantum-critical point

Bernd Wolf[a,1], Yeekin Tsui[a,2], Deepshikha Jaiswal-Nagar[a], Ulrich Tutsch[a], Andreas Honecker[b], Katarina Remović-Langer[a], Georg Hofmann[a], Andrei Prokofiev[c], Wolf Assmus[a], Guido Donath[d], and Michael Lang[a]

[a]Physics Institute, Goethe-University, Max-von-Laue Strasse 1, 60438 Frankfurt(M), Germany;

[b]Institute for Theoretical Physics, Georg-August-University Göttingen, Friedrich-Hund-Platz 1, 37077 Göttingen, Germany;

[c]Institute of Solid State Physics, Vienna University of Technology, Wiedner Hauptstrasse 8-10, 1040 Vienna, Austria;

[d]Max-Planck-Institut for Chemical Physics of Solids, Nöthnitzer Strasse 40, 01187 Dresden, Germany;

[1]To whom correspondence may be addressed. E-mail: wolf@physik.uni-frankfurt.de

[2]Present address: Department of Physics, Durham University, United Kingdom

Classification: Physical Sciences/Physics

# Abstract

The presence of a quantum critical point (QCP) can significantly affect the thermodynamic properties of a material at finite temperatures $T$. This is reflected, e.g., in the entropy landscape $S(T, r)$ in the vicinity of a QCP, yielding particularly strong variations for varying the tuning parameter $r$ such as pressure or magnetic field $B$. Here we report on the determination of the critical enhancement of $\partial S/\partial B$ near a $B$-induced QCP via absolute measurements of the magnetocaloric effect (MCE), $(\partial T/\partial B)_S$, and demonstrate that the accumulation of entropy around the QCP can be used for efficient low-temperature magnetic cooling. Our proof of principle is based on measurements and theoretical calculations of the MCE and the cooling performance for a $Cu^{2+}$-containing coordination polymer, which is a very good realization of a spin-½ antiferromagnetic Heisenberg chain – one of the simplest quantum-critical systems.

**Introduction**

The magnetocaloric effect (MCE), i.e. a temperature change in response to an adiabatic change of the magnetic field, has been widely used for refrigeration. Although, up until now applications have focussed on cryogenic temperatures (1-3), possible extensions to room temperature have been discussed (4). The MCE is an intrinsic property of all magnetic materials in which the entropy $S$ changes with magnetic field $B$. Paramagnetic salts have been the materials of choice for low-temperature refrigeration (1), including space applications (5-7), with an area of operation ranging from about one or two degrees Kelvin down to some hundredths or even thousandths degree Kelvin. Owing to their large $\Delta S/\Delta B$ values, the ease of operation, and the applicability under microgravity conditions, paramagnets have matured to a valuable alternative to $^3$He-$^4$He dilution refrigerators, the standard cooling technology for reaching sub-Kelvin temperatures.

A large MCE also characterizes a distinctly different class of materials, where the low-temperature properties are governed by pronounced quantum many-body effects. These materials exhibit a $B$-induced quantum-critical point (QCP) – a zero-temperature phase transition –, and the MCE has been used to study their quantum criticality (8-12). The aim of the present work is to provide an accurate determination of the enhanced MCE upon approaching a $B$-induced QCP both as a function of $B$ and $T$ and to explore the potential of this effect for magnetic cooling.

Materials in the vicinity of a QCP have been of particular current interest, as their properties reflect critical behavior arising from quantum fluctuations instead of thermal fluctuations which govern classical critical points (13). Prominent examples of findings made here include the intriguing low-temperature behaviors encountered in some heavy-fermion metals (14, 15) or magnetic insulators (16,17) and the occurrence of new quantum phases near a QCP (18, 19, 10). Generally, a QCP is reached upon tuning an external parameter $r$ such as pressure or magnetic field to a critical value. Although the critical point is inaccessible by experiment, its presence can affect the material's properties at finite temperature significantly. The thermodynamic properties are expected to show anomalous power-laws as a function of temperature and, even more spectacular, to exhibit an extraordinarily high sensitiveness on these tuning parameters (20, 21). Upon approaching a pressure ($p$)-induced QCP, for example, the thermal expansion coefficient $\alpha \propto \partial S/\partial p$ is more singular than the specific heat $C = T \cdot \partial S/\partial T$, giving rise to a diverging Grüneisen ratio $\Gamma_p \propto \alpha/C$ (20, 22). Likewise, for a $B$-induced QCP, a diverging $\Gamma_B = -C^{-1} \cdot (\partial S/\partial B)_T$ is expected (20, 21, 23) and has recently been reported (12). Unlike $\Gamma_p$, however, where experimental access, apart from measurements of the critical contributions of the two quantities $\alpha$ and $C$, is very difficult, $\Gamma_B$ can be determined directly via the magnetocaloric effect (MCE) by probing temperature changes of the material in response to changes of the magnetic field $B$ under adiabatic conditions,

$$\Gamma_B = \frac{1}{T}\left(\frac{\partial T}{\partial B}\right)_S. \quad (1)$$

In this work, we directly measure the critical enhancement of $\Gamma_B$ near a $B$-induced QCP and demonstrate that this effect can be used for magnetic refrigeration over an extended range of temperatures. Our results indicate that quantum critical materials open up new possibilities for realizing very efficient and flexible low-temperature coolants.

**The model substance**

For the proof of principle, it is helpful to focus on simple model substances, characterized by a small number of material parameters which are well under control. One of the simplest quantum-critical systems, where an enhanced MCE (23, 21) and a potential use for magnetic cooling (24) have been predicted, is the uniform spin-½ antiferromagnetic Heisenberg chain (AFHC), described by

$$H = J \sum_i \vec{S}_i \cdot \vec{S}_{i+1}. \qquad (2)$$

This model, and its highly non-trivial physics, contains a single parameter $J$ – the Heisenberg exchange interaction – by which nearest-neighbor spins interact along one crystallographic direction. In fact, by exploring the quasi-1D spin-½ Heisenberg antiferromagnet $KCuF_3$, it has been demonstrated that quantum-critical Luttinger liquid (LL) behavior governs the material's properties over wide ranges in temperature and energy (17). The AFHC remains in the LL quantum-critical state also in magnetic fields (25) up to the saturation field $B_s$, given by $g\mu_B B_s = 2J$, with $g$ the spectroscopic $g$ factor and $\mu_B$ the Bohr magneton. Since the fully polarized state above $B_s$ is a new eigenstate of the system, different from that of the LL, $B_s$ marks the endpoint of a quantum-critical line in the $B$-$T$ plane.

For exploring the MCE around $B_s$, we use a copper-containing coordination polymer $[Cu(\mu\text{-}C_2O_4)(4\text{-aminopyridine})_2(H_2O)]_n$ (abbreviated to CuP henceforth), a good realization of a spin-½ AFHC, and compare the results with model calculations for the idealized system. In CuP, first synthesized by Castillo *et al.* (26), $Cu^{2+}$ (spin-½) ions, bridged by oxalate molecules ($C_2O_4$), form chains along the crystallographic $c$-axis (26, 27). High-quality single crystals of CuP, made recently available by a slow-diffusion technique (27), were used for the present investigations (see *SI Text I*). The magnetic susceptibility of these single crystals for $T \geq 0.055$ K, are well described by the model of a spin-½ AFHC (28) with an *intra*-chain coupling $J/k_B = (3.2 \pm 0.1)$ K (27).

In order to fully characterize the magneto-thermal properties of CuP, measurements of the specific heat $C_B(T)$ and low-temperature magnetic susceptibility $\chi_T(B)$ have been carried out (Fig. 1). The excellent agreement of these results with model calculations (solid lines in Fig. 1) (see *SI Text II*) confirms the assertion (26, 27) that CuP is a very good realization of a uniform spin-½ AFHC. Compared to other model substances of this kind, such as copper pyrazin dinitrate (29), however, CuP excels by its moderate size of $J$ and the correspondingly small saturation field of 4.09 T (for $B \parallel b$), enabling the MCE to be studied in the relevant field range below about twice the saturation field by using standard laboratory magnets.

**Absolute measurements of the magnetocaloric effect and comparison with theory**

For a quantitative determination of $\Gamma_B$, a novel step-like measuring technique was employed (see *SI Text III and IV*) ensuring, to good approximation, adiabatic conditions. The $\Gamma_B$ data obtained this way for two different temperatures exhibit negative values for $B < B_s$ (Fig. 2A), implying that here cooling is achieved through magnetization! Upon increasing the field, $\Gamma_B$ passes through a weak minimum, changes sign and adopts a pronounced maximum at fields somewhat above 4 T. The large positive $\Gamma_B$ values here indicate that in this $B$-$T$ range, a pronounced cooling effect is obtained through demagnetization. The sign change in $\Gamma_B \propto \partial S/\partial B$ within a narrow $B$-interval centered near $B_s$ implies the presence of a distinct maximum in $S_T(B)$ – a clear signature of a nearby $B$-induced QCP (21). It reflects the accumulation of entropy due to the competing ground states separated by the QCP. The data taken at 0.32 K show extrema in $\Gamma_B$ which are distinctly sharper and more pronounced as compared to the signatures in the 0.97 K data, suggesting incipient divergencies of $\Gamma_B$ on both sides of $B_s$. Theoretical calculations of $\Gamma_B$ have been performed for the ideal spin-½ AFHC with $J/k_B = 3.2$ K (solid lines in Fig. 2A). The model curves capture the essential features of the experimental results and even provide a good quantitative description for the data taken at 0.97 K. However, some systematic deviations become evident which increase with decreasing temperature: the extrema in $\Gamma_B$ as a function of field at fixed temperature (Fig. 2A) are less strongly pronounced compared to theoretical results, and the experimental data exceed the theory curves at high fields. To follow the evolution of the extrema as a function of temperature, measurements at $B$ = const. have been performed (insert of Fig. 2B). The deviations from the model curves suggest that by the application of a magnetic field, the divergence in $\Gamma_B(T \to 0)$ becomes truncated by the occurrence of a small energy scale, accompanied by a shift of entropy to higher fields. A plausible explanation for these observations would be the opening of a small field-induced gap. It may result, e.g., from a finite Dzyaloshinskii-Moriya (DM) interaction (31, 32) permitted by the lack of a centre of inversion symmetry in CuP (26). In fact, indications for a finite DM interaction can be inferred from measurements performed under different field orientations (†). In addition, part of the deviations from the model curves at low temperatures can also be due to the presence of weak inter-chain interactions and the accompanied dimensional crossover, inevitable in any three-dimensional material. It may also in part reflect the influence of an enhanced thermal boundary resistance between sample and thermometer due to an enhanced spin-phonon interaction caused by a nearby phase transition (33).

---

† Besides lacking inversion symmetry, CuP has a two-fold rotation axis parallel to the *b*-axis which constrains the DM-vector **D** to lie within the *ac* plane. A finite DM interaction in CuP is consistent with the results of magnetic cooling experiments for two different orientations of the magnetic field with respect to **D**: the accessible lowest temperature $T_f$ raises by about $(10 \pm 2)$ % for $B \parallel c$ as compared to $B \parallel b$.

**Magnetic cooling and performance characteristics**

Despite these deviations from the idealized system, the anomalous MCE (Fig. 2A) highlights distinct quantum-critical behavior in CuP over extended ranges of temperature and magnetic field. The strongly enhanced values of $|\Gamma_B|$, implying large temperature changes in response to small field variations, suggest that the system could be suitable for low-temperature magnetic refrigeration. To fathom out its potential as a coolant, demagnetization experiments were carried out on CuP under near adiabatic conditions (see *SI Text III*) while simultaneously recording the sample temperature $T_s$ (Fig. 2B). While the cooling process approaches an in-$B$ linear behavior at higher temperature near ($T_i$, $B_i$) – such is seen in simple paramagnets where $T_s \propto B$ –, it becomes superlinear upon decreasing the temperature. This enhanced cooling effect, which is in accordance with the model calculations for the ideal system (broken lines in Fig. 2B), is a direct manifestation of quantum criticality. Upon further cooling, $T_s(B)$ passes through a rounded minimum, assigned $T_0$, for $B$ close to $B_s$. The quantitative deviations from the theory curves, in particular an experimentally revealed $T_0$ staying clearly above the theoretically expected value, can be attributed only in part to non-ideal adiabatic conditions in the cooling experiments. Estimates of the effect of the parasitic heat flow onto the sample (see *SI Text V*) show that for an improved thermal isolation, $T_0$ can be reduced from 179 mK to about 132 mK (solid line in Fig. 2B). This implies that the main source for the deviations from the theoretical expectations lies in the presence of the above-mentioned perturbing interactions in CuP. The good agreement with the theory curves at higher temperatures, where these interactions are irrelevant, however, implies that for a better realization of the spin-½ AFHC, cooling to much lower temperatures should be possible.

In order to assess the principle limits of cooling near a QCP and its potential for applications, we compare some performance characteristics for the ideal spin-½ AFHC with those of two state-of-the-art adiabatic demagnetization refrigerator (ADR) materials (Fig. 3). These are $CrK(SO_4)_2 \cdot 12H_2O$ (chrome potassium alum CPA) and $FeNH_4(SO_4)_2 \cdot 12H_2O$ (ferric ammonium alum FAA), which cover the typical temperature range of applications for paramagnetic salts and, due to their high efficiency, are also used in space applications (5-7). Representative parameters characterizing the performance of an ADR material include (i) the operating range, in particular its lower bound $T_{min}$. Standard ADR systems operate predominantly in a narrow temperature window near $T_{min} \approx$ 30-60 mK, while extensions down to only a few millikelvin are possible. Such low temperatures are otherwise accessible only by using elaborated $^3$He-$^4$He dilution refrigerators, which, however, are much bigger compared to ADRs and cannot operate in a microgravity environment. Another important parameter is (ii) the "hold time" of the coolant, which is inversely proportional to its cooling power $\dot{Q}$. This quantity measures the ability of the refrigerant to absorb heat without warming up too rapidly. For some applications, (iii) the efficiency can be of importance. This includes the cooling capability per unit mass of refrigerant material and the ratio $\Delta Q_c/\Delta Q_m$. Here $\Delta Q_c$ is the heat the material can absorb after demagnetization to the final field $B_f$, and $\Delta Q_m$ the heat of magnetization released to a precooling stage held at a temperature $T_i$, the initial temperature of the magnetic cooling process. The efficiency can be an issue for modern multi-stage single-shot or continuous ADRs, such as the ones used in space (5), where the entire system has to be optimized with regard to precooling requirements and weight.

In principle, the cooling performance of a material is determined by the low-energy sector of its magnetic excitation spectrum, reflected in the low-temperature specific heat $C(T)$, and its variation with magnetic field. For paramagnetic salts, $C(T)$ at $B = 0$ is of Schottky type (see, e.g., ref. 33) due to an energy-level splitting arising from residual magnetic interactions. The resulting $C(T)$ maximum actually enables the material to be used as a coolant as it ensures a certain amount of cooling power during demagnetization at low temperatures. At the same time, the $1/T^2$-like decrement of the $C(T)$ anomaly at high temperatures and the accompanied rapid reduction of the hold time (Fig. 3) constrain the operating range to a narrow temperature window around the position of the maximum. Thus, ADRs based on paramagnets are particularly well suited for applications within a narrow temperature range of operation.

This is different for the spin-½ AFHC near $B_s$: due to its peculiar excitation spectrum (28), i.e. the abundance of low-energy excitations above the QCP, (in principle) arbitrarily small values of $T_{min}$ can be reached, while keeping the cooling power considerably large. At the same time, due to the extraordinarily large specific heat way above the QCP, with a peak centered at a temperature around half of $J/k_B$ (Fig. 1), the system affords extended hold times also for temperatures largely exceeding the maximum position (Fig. 3). Thus, materials near a $B$-induced QCP can be an excellent alternative to paramagnets for those applications where the temperature has to be varied over an extended range. This includes the possibility to reach very low temperatures, albeit with reduced hold times. In addition, the present quantum critical system, despite its low spin value, can absorb an amount of energy $\Delta Q_c$ which is almost comparable to that of the spin-3/2 (CPA) and spin-5/2 (FAA) paramagnets (Fig. 4). The quantum critical system excels, however, by its high efficiency $\Delta Q_c/\Delta Q_m$, which exceeds the corresponding numbers for the paramagnets by a factor 2-3 (Fig. 4). This is due to the system's ability to absorb energy even at temperatures as high as $T_i$, in contrast to the paramagnets, where the absorption essentially occurs at low temperatures $T < T_i$ (Fig. 4). Besides these performance characteristics, the applicability of a material as a coolant may also depend on other material-specific features such as its thermal conductivity and the possibility to realize a good thermal contact to the body to be cooled. In terms of heat transport, the spin-½ AFHC near the *B*-induced QCP is particularly favorable as it shows a comparatively large spin thermal conductivity along the spin chains, which even dominates the material's thermal transport at low temperatures, see, e.g. ref. 35. In contrast, a small thermal conductivity and a weak thermal contact between the coolant and another body are principal concerns for paramagnetic ADR materials (33). Here, problems can arise due to the extreme hydration of these materials, required to reduce the mutual interactions between the magnetic centres, and the materials' sensitivity to decompose if water is allowed to evaporate through imperfections in the materials' housing.

**Conclusions and outlook**

We have demonstrated that the strong enhancement of the magnetocaloric effect, arising from quantum fluctuations near a *B*-induced quantum-critical point, can be used for realizing an efficient and flexible magnetic cooling with a good cooling performance over an extended range of temperatures. While the spin-½ AFHC, thanks to its simplicity, has enabled the provision of a proof-of-principle demonstration, extensions of this concept are obvious and may guide the search for materials with further improved cooling performance. On the one hand, this includes low-dimensional spin systems with geometric frustration, such as the saw-tooth chain discussed in ref. 23. In contrast to the uniform spin-½ AFHC discussed here, where the enhanced MCE results from large relative changes of the entropy across $B_s$, while $S$ vanishes for $T \to 0$, certain geometrically frustrated spin configurations in one dimension support a significant zero-temperature entropy (36). As a consequence, in these systems there is a large absolute variation of entropy across $B_s$ and hence a large $\partial T/\partial B$ down to, in principle, arbitrarily low temperatures, limited by weak residual interactions, such as dipole-dipole interactions. Consequently, a further enhancement of the MCE can be expected for those frustrated magnetic systems. The combination of quantum criticality with (i) low dimensionality, assuring a high density of low-energy excitations, and (ii) geometric frustration, opens up a promising route to search for materials with further enhanced MCE. Likewise, as a large spin entropy can cause a large thermopower in conducting materials (37), conducting magnets driven close to a field-induced QCP would be a very interesting class of materials for the search for thermoelectric (Peltier) cooling systems with high efficiency. Finally, the concept of using quantum criticality for magnetic cooling might also be applicable to cold atom experiments and help there to realize low-temperature magnetic and superfluid phases.

**Materials and methods**

The specific heat was measured as a function of temperature by employing a compensated heat-pulse technique in combination with a $^3$He-$^4$He dilution refrigerator (see *SI Text I*). For the magnetic susceptibility measurements a state-of-the-art compensated-coil ac-susceptometer adapted to a top-loading $^3$He-$^4$He dilution refrigerator was used (see *SI Text I*). The MCE was determined by employing a specially-designed calorimeter (see *SI Text III*). For the magnetic cooling experiments, the setup was modified to ensure quasi-adiabatic conditions (see *SI Text IV*).

**Acknowledgments**

A. H. acknowledges financial support by the Deutsche Forschungsgemeinschaft (DFG) via a Heisenberg fellowship. Work at Goethe-University Frankfurt was supported by the DFG in the consortium SFB/TR49.

[37] Wang Y, Rogado N S, Cava R J, Ong N P (2003) Spin entropy as the likely source of enhanced thermopower in $Na_xCo_2O_4$. *Nature* 423:425-428.

**Figures 1-4:**

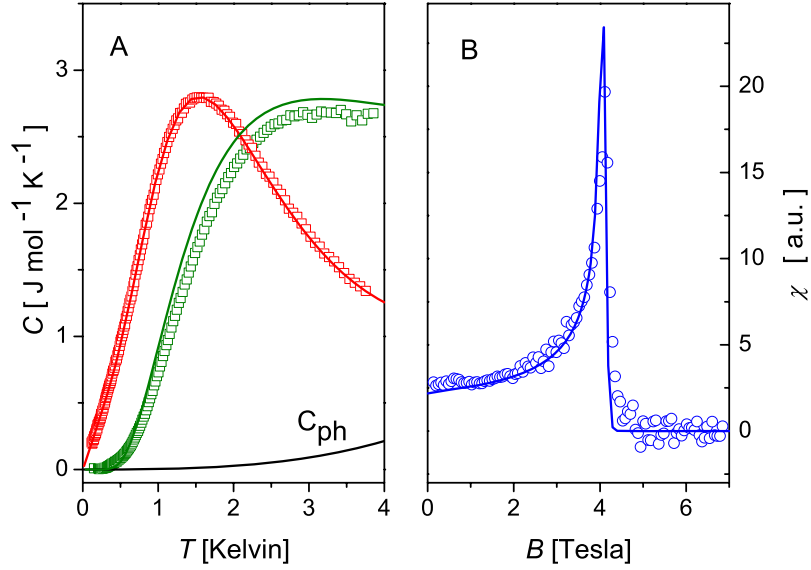

**Fig. 1**. **(A)** Specific heat $C_B(T)$ measured in a constant magnetic field of $B = 0.15$ T (red symbols) and 7.14 T (green symbols). While the data at $B = 0.15$ T $\ll B_s$ (saturation field) are characterized by an in-$T$ linear contribution $C/T \propto J^{-1}$ for $T \ll J/k_B$ (28), those at $B = 7.14$ T $> B_s$ show an activated behavior, reflecting the opening of an excitation gap in the field-induced ferromagnetic state. The red solid line is the result of Bethe Ansatz calculations (BAC) (28) for a spin-½ AFHC with $J/k_B = 3.2$ K, $g = 2.087$, $B/B_s = 0.0329$, representing the magnetic contribution $C_{mag}$, in addition to a small phonon contribution $C_{ph} = \beta \cdot T^3$ (black solid line). The coefficient $\beta$, the only adjustable parameter, was determined from a least-squares fit of $C = C_{ph} + C_{mag}$ to the experimental data taken at $B = 0.15$ T. The green solid line corresponds to $C_{ph} + C_{mag}$, with $C_{mag}$ obtained from BAC for $B/B_s = 1.564$, corresponding to $B = 7.14$ T. The excellent agreement of the $C_B(T)$ data with the model calculations at small fields and the moderate deviations at 7.14 T indicate the presence of additional field-dependent terms in the material's Hamiltonian which become relevant at high fields. **(B)** Blue spheres represent the magnetic susceptibility $\chi = \partial M/\partial B$, taken on single-crystalline material immersed in the $^3$He-$^4$He mixture kept at a constant temperature $T = 0.055$ K with $B \parallel b$-axis. The field was swept at a low rate of 0.1 T/min, ensuring constant-temperature conditions. The growth of $\chi$ upon increasing the field, the sharp peak at 4.09 T and the disappearance of $\chi$ at higher fields reflect the transition from the antiferromagnetic ($B < B_s$) to the polarized ferromagnetic state ($B > B_s$) at $B_s = 4.09$ T. The solid line represents a parameter-free description of the data based on the above-mentioned BAC for $J/k_B = 3.2$ K and $g = 2.33$ (27).

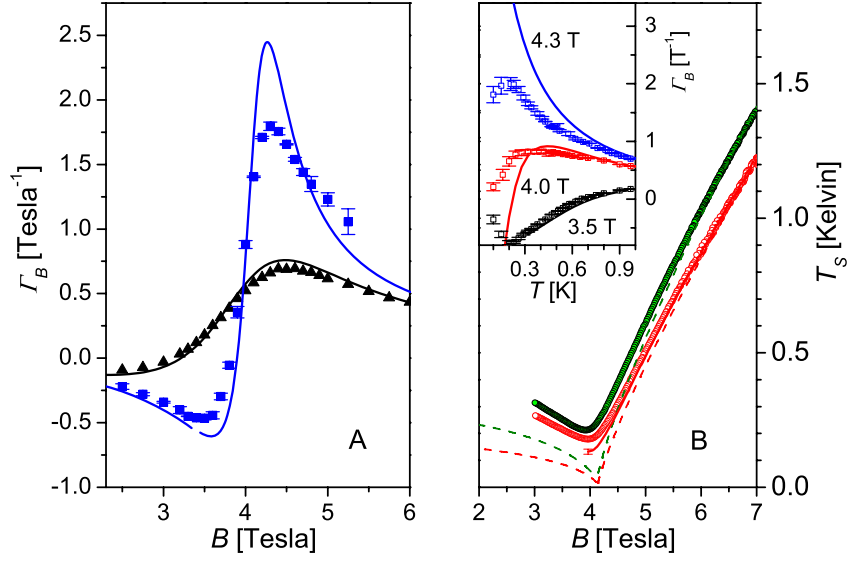

**Fig. 2.** (**A**) Variation of $\varGamma_B$, approximated by $\varGamma_B \approx T^{-1} \cdot (\Delta T/\Delta B)_{S \approx \text{const}}$ (see *SI Text IV*), with field at $T = 0.32$ K (blue squares) and 0.97 K (black triangles) for $B \parallel b$-axis. For $\varGamma_B$ data at an intermediate temperature $T = 0.47$ K, see ref. 11. Solid lines are the results of Quantum-Monte-Carlo (QMC) simulations and exact diagonalization of a finite-size lattice of the spin-½ AFHC for the corresponding temperatures (same color code) by using $J/k_B = 3.2$ K. A $g$ value of 2.28 has been used in the calculations to account for a small misalignment of the crystals. Error bars (below symbol size for the data taken at 0.97 K) of experimental data correspond to statistical error. The systematic error, mainly originating from the internal time constant of the resistance bridge in combination with the sweep time applied, is estimated to be of the order of 1 %. (**B**) Sample temperature (symbols) measured by demagnetizing a collection of CuP single crystals ($B \parallel b$-axis) of total mass of 25.23 mg under near adiabatic conditions (see *SI Text III*). The initial parameters were set to $B_i = 7$ T and $T_i = 1.40$ K (green symbols) and 1.21 K (red symbols). The field was swept with a rate $\Delta B/\Delta t = -0.3$ T/min for $B \geq 6$ T and -0.5 T/min for $B < 6$ T. The red solid line represents the experimental data (for $T_i = 1.21$ K) corrected for the effect of a parasitic heat flow (see *SI Text V*). Broken lines are those ideal isentropes, derived from the exact result for the entropy of the spin-½ AFHC (30), which correspond to the same initial conditions ($B_i$, $T_i$). The **insert in (B)** shows the temperature dependence of $\varGamma_B$, approximated by $\varGamma_B \approx T^{-1} \cdot (\Delta T/\Delta B)_{S \approx \text{const}}$ (see *SI Text IV*), at constant fields below ($B = 3.5$ and 4.0 T) and above (4.3 T) $B_s$. The deviations of the experimental data from the model calculations for the ideal system (solid lines, same color code as used for the experimental data) grow with decreasing the temperature. The data reveal an anomaly around 0.22 K, the position of which shows no significant field dependence within the field range investigated. This feature is likely to be related to the opening of a field-induced gap in the magnetic excitation spectrum.

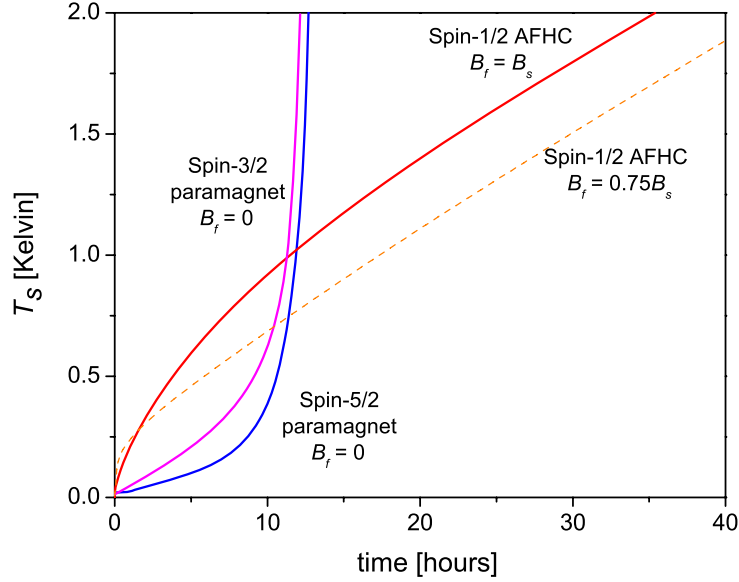

**Fig. 3.** Increase of the sample temperature $T_s$ with time after demagnetization to $B = B_f$ calculated for a heat load of 5 μW applied under adiabatic conditions to 100 gram substance of a spin-1/2 AFHC with $J/k_B = 3.2$ K (red solid line), a spin-3/2 (CrK(SO$_4$)$_2$·12H$_2$O in short CPA) (magenta solid line) and a spin-5/2 (Fe(NH$_4$)(SO$_4$)$_2$·12H$_2$O in short FAA) (blue solid line) paramagnetic salt (see *SI Text VI*). For the paramagnets at $B_f = 0$, base temperatures of $T_{min} \approx 0.01$-$0.015$ K (CPA) and $\approx 0.04$-$0.05$ K (FAA) can be reached, which lie close to the transition temperature to long-range antiferromagnetic order at $T_N = 0.01$ K (CPA) and 0.04 K (FAA). In contrast, for the spin-1/2 AFHC, cooling to an (in principle) arbitrarily small $T_{min}$ is possible. In practice, however, a finite $T_{min}$ will result from the presence of small perturbing interactions such as weak inter-chain couplings, dipole-dipole interactions or single-ion anisotropies. The calculations for $T_s$(time) are based on computed (spin-1/2 AFHC) and measured (CPA and FAA) (33) specific heat data. The paramagnets exhibit long hold times only in a narrow temperature range around $T_{min}$, defined by the position of their Schottky-type $C(T,B=0)$ maximum. For temperatures somewhat above $T_{min}$, $T_s$ rapidly increases with exposure time $t$, approximately as $t^2$ for FAA. This contrasts with the spin-1/2 AFHC, where $T_s$ increases only moderately with time as $t^{2/3}$. A further flattening of $T_s(t)$, at not too low temperatures, can be achieved for $B_f < B_s$; see, e.g. the warming curve for $B_f = 0.75B_s$ (broken orange line). Crossing $B_s$ upon demagnetization from $(B_i, T_i) = (7$ T, 1.3 K) is, however, accompanied by a moderate warming to $T_{min} \approx 0.112$ mK.

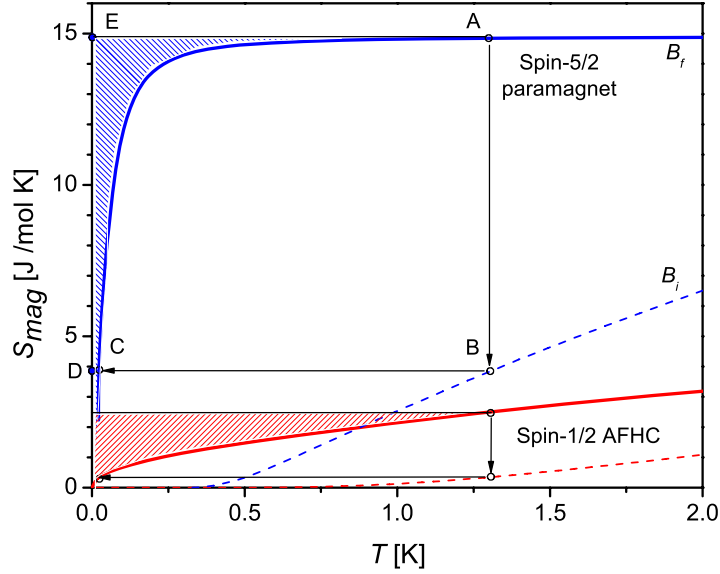

**Fig. 4.** Calculated molar magnetic entropy $S_{mag}(T, B=\text{const.})$ as a function of temperature of the spin-1/2 AFHC with $J/k_B = 3.2$ K for an initial field of $B_i = 7.14$ T (red broken line) and a final field $B_f = B_s = 4.09$ T (red solid line). For comparison, $S_{mag}(T, B=\text{const.})$ of the spin-5/2 paramagnetic salt Fe(NH$_4$)(SO$_4$)$_2$·12H$_2$O is shown for $B_i = 2$ T (blue broken line) and $B_f = 0$ (blue solid line); the data were taken from refs. 33 and 34. In the cooling process, the materials are first isothermally magnetized (path AB for example), and then, after thermal isolation, adiabatically demagnetized (path BC) to the final temperature $T_f$. For $T_i = 1.3$ K, for example, adiabatic demagnetization from 7.14 T to 4.09 T for the spin-1/2 AFHC results in $T_f = 0.021$ K, while demagnetization from 2 T to zero field for the spin-5/2 paramagnet leads to $T_f = 0.026$ K. The systems warm up along their entropy curves at the final demagnetization field $S_{mag}(T, B=B_f)$. The heat of magnetization at $T = T_i$, $\Delta Q_m = T_i \cdot [S_{mag}(B_f, T_i) - S_{mag}(B_i, T_i)]$, and the heat that the material is able to absorb after adiabatic demagnetization, $\Delta Q_c = \int_{T_f}^{\infty} T \cdot (\partial S_{mag}/\partial T)_{B_f} dT$, can be read off the figure. For the spin-5/2 paramagnet (PM) Fe(NH$_4$)(SO$_4$)$_2$·12H$_2$O, for example, $\Delta Q_m = 14.3$ J/mole is given by the area of the rectangle ABDE, while $\Delta Q_c = 1.22$ J/mole corresponds to the hatched area in that rectangle. The efficiency factor $\Delta Q_c/\Delta Q_m$ in the temperature range indicated amounts to 26 % for the spin-1/2 AFHC as compared to only 9 % for the spin-5/2 paramagnet. For the spin-3/2 system CrK(SO$_4$)$_2$·12H$_2$O (see *SI Text VII*) one finds $\Delta Q_m = 10.01$ J/mole, $\Delta Q_c = 1.06$ J/mole and $\Delta Q_c/\Delta Q_m$ of 11 %.